\documentclass[12pt]{article}

\begin{document}
\title{{\bf Particle dynamics on hyperboloid\\
 and unitary representation\\ of $SO(1,N)$ group}}
\author{George Jorjadze$^\dag$ and 
W{\l}odzimierz Piechocki$^\ddag$ \\
{\small $^\dag$Razmadze Mathematical Institute, 
Tbilisi, Georgia }\\
{\small $^\ddag$So{\l}tan Institute for Nuclear Studies,  
Warsaw, Poland }}
\date{}
\maketitle

\begin{abstract}
\noindent
We analyze particle dynamics on $N$ dimensional
one-sheet hyperboloid embedded in $N+1$ dimensional
Minkowski space. 
 The dynamical integrals constructed by $SO_\uparrow (1,N)$ symmetry
of spacetime are used for the 
gauge-invariant Hamiltonian reduction. 
 The physical phase-space parametrizes the set of all 
classical trajectories on the hyperboloid. 
In quantum case the operator
ordering problem for the symmetry generators is solved by transformation
to asymptotic variables. Canonical quantization leads to 
unitary irreducible representation of $SO_\uparrow (1,N)$ group
on Hilbert space $L^2(S^{N-1})$. 
\end{abstract}

\section{Introduction}

In recent papers [1-3] we investigated classical and quantum dynamics
of a relativistic particle on 2-dimensional Lorentzian manifold with constant
curvature. Locally, such manifolds (with the same curvature)
are isometric, they are described by the Liouville equation
and have the symmetry associated with $sl(2,R)$ algebra.
However, the global properties of these manifolds can be different.

For the classical description we have used the following scheme:

\noindent
Reparametrization invariant Lagrangian of the system has a symmetry 
of spacetime.
The dynamical integrals constructed by this symmetry define the physical
phase-space $\Gamma_{ph}$, which is specified by the admissible
values of dynamical integrals on the constraint surface.
The physical phase-space fixes the Casimir number of $sl(2,R)$ algebra
and gives $C=m^2r^2$, where $m$ 
is particle mass and $r$ is radius of hyperboloid.
Independent physical variables are used to construct the symplectic form
on $\Gamma_{ph}$ [4]. Locally, the physical-phase space and
spacetime have the same symmetry, which is expressed by
 different structures (symplectic
and metric, respectively). The physical phase-space parametrizes
the set of all classical trajectories in spacetime. 
 In this way we find the correlation
between global properties of spacetime and physical phase-space.

The symplectic structure on $\Gamma_{ph}$ is used for the quantization of the
reduced system. For the operators of the dynamical integrals one has
ambiguity connected with operator ordering. It is shown that this problem
can be solved in case $\Gamma_{ph}$ has global 
$SO_\uparrow (1,2)$ symmetry.

The 2d one-sheet hyperboloid embedded in 3d
Minkowski space is an example of a constant curvature manifold with
 global $SO_\uparrow (1,2)$ symmetry. 
 In this case the physical phase space has the 
same global symmetry. The corresponding quantum theory describes 
 unitary irreducible representation of
$SO_\uparrow (1,2)$ group. 

The present paper is a generalization of these results for  
particle dynamics on $N$ dimensional hyperboloid, embedded in
$N+1$ dimensional Minkowski space, for arbitrary $N$.
Such manifold has constant curvature [5] and 
global $SO_\uparrow (1,N)$ symmetry, which plays the same role in this curved
spacetime as the Poincare group in flat Minkowski space.

In classical case we follow the scheme just described for the case 
$N=2$. 
To have a covariant description we use the coordinates of
$N+1$ dimensional Minkowski space, 
which leads to the additional constraints.

In quantum theory we have the problem of operator ordering for the
symmetry generators. Generalizing the method of [1], we find unitary 
irreducible representation of the symmetry group. As far as we know
it is a new representation of $SO_\uparrow (1,N)$ group.

\section{Classical dynamics on hyperboloid}  

Let $x^\mu$ $(\mu = 0,1,...,N)$ be the coordinates on $N+1$
dimensional Minkowski 
space with the metric tensor $\eta_{\mu\nu}=diag(+,-,...,-)$.
A one-sheet hyperboloid $\mathbf{H}_N$ is defined by
\begin{equation}
\b{x}^2+r^2=0~~~~~~~~~~~(\b{x}^2 :=\eta_{\mu\nu}x^\mu x^\nu ), 
\end{equation}
where $r>0$ is a fixed parameter.
The induced metric on $\mathbf{H}_N$ has the Lorentzian signature.
 The topology of $\mathbf{H}_N$ is 
$R^1\times S^{N-1}$ and
$\mathbf{H}_N$ can be considered as a model of `universe'
with a compact ($S^{N-1}$) space.

We describe the dynamics of a relativistic particle  on $\mathbf{H}_N$
by the action 
\begin{equation}
S=\int d\tau [-m\sqrt{\dot{\b{x}}^2} +\lambda (\b{x}^2 +r^2)],
\end{equation}
where $\tau$ is an evolution parameter, $\lambda$
plays a role of Lagrange multiplier, $m$ is particle mass and
$ \dot{\b{x}}^2:=
\eta_{\mu\nu}\dot{x}^\mu\dot{x}^\nu$ ($\dot{x}^\mu := 
dx^\mu/d\tau$). The coordinate $x^0$ is associated with time
and we assume that 
$\dot x^0>0$.

The action (2) is invariant under the Lorentz transformations 
\begin{equation}
x^\mu \rightarrow \Lambda^\mu\,_\nu\, x^\nu,~~~~~\Lambda^\mu\,_\nu\in 
SO_\uparrow(1,N)
\end{equation}
and the corresponding dynamical integrals constructed by the
 Noether theorem have the form
\begin{equation}
M_{\mu\nu} =p_{\mu}x_\nu -p_{\nu}x_\mu , 
\end{equation}
where $p_{\mu}$ are the canonical momenta. The variables ($x^\mu,
p_\mu$) define $2(N+1)$ 
dimensional extended phase-space 
$\Gamma_{e}$.

In the Hamiltonian formulation the Dirac procedure [6]
leads to the following constraints
\begin{equation}
\Phi_1 = \b{x}^2+ r^2 =0,~~~~\Phi_2 =\b{p}^2 -m^2 =0,
~~~~\Phi_2 =\b{p}\b{x}=0,  
\end{equation}
where $\b{p}^2 :=\eta^{\mu\nu}p_\mu p_\nu$, $\b{p}\b{x}:=p_\mu x^\mu$.
These constraints fix the Casimir number 
\begin{equation}
C :=\frac{1}{2} M_{\mu\nu}M^{\mu\nu} =m^2r^2 . 
\end{equation}

The constraint surface $\Gamma_{c}$,
given by (5), is $2N-1$ dimensional submanifold of $\Gamma_{e}$.
By Eq.(4) we have the map $F$ from $\Gamma_{e}$ to the space 
of dynamical integrals $\Gamma_{d}$. 
The physical phase-space $\Gamma_{ph}$ is a submanifold of $\Gamma_{d}$
defined by $\Gamma_{ph}=F(\Gamma_{c})$.
To specify $\Gamma_{ph}$
we introduce the variables
\begin{equation}
\xi_n :=\frac{J_n}{J},~~~~~~~\eta_n:=\frac{J_mM_{mn}}{J},~~~~~(m,n =1,...,N), 
\end{equation}
where $M_{mn}$ are generators of space rotations in $N+1$ dimensional
Minkowski space, $J_n:=M_{n0}$ are generators for boosts and 
$J:=\sqrt{J_kJ_k}$. 

Due to (7) we have
\begin{equation}
\xi_k\xi_k=1~~~~~~~~\mbox{and}~~~~~~~~
\xi_k\eta_k=0.
\end{equation}
Therefore, Eq.(7) gives a map from $\Gamma_{d}$ to $TS^{N-1}$. 
The  map (7) is invertible on $\Gamma_{ph}$,
since from (4) -- (8) we obtain
\begin{equation}
M_{mn} =\xi_m\eta_n -\xi_n\eta_m,~~~~ J_n =\xi_n\sqrt{\kappa^2 +\eta^2},
\end{equation}
where $\kappa:=mr$ and $\eta :=\sqrt{\eta_k\eta_k}$.

Thus, we conclude that the manifold $TS^{N-1}$, given by (8), 
represents the physical phase-space
$\Gamma_{ph}$.

To find the symplectic structure on $\Gamma_{ph}$ we use the method 
based on calculation of a reduced 1-form $p_\mu dx^\mu$ 
in physical variables (see [4], [7]). 
For this purpose we parametrize the constraint surface
(5) by the physical variables $\xi_n, \eta_n$ and the parameter
$s=-\tanh^{-1}(mx_0/rp_0)$. From (4), (5) and (7) we get
\begin{eqnarray}
p_0 =-\frac{1}{r}\sqrt{\kappa^2+\eta^2}\cosh s,~~~~~~~~~~
x_0 =\frac{1}{m}\sqrt{\kappa^2+\eta^2}\sinh s, \\ \nonumber
p_n=-\frac{\eta_n}{r}\cosh s -m\xi_n\sinh s,~~~~
x_n=\frac{\eta_n}{m}\sinh s + r\xi_n\cosh s
\end{eqnarray}
(we take into account that $p_0<0$, since $\dot{x}^0 >0$).

In the parametrization (10) the canonical 1-form $\Theta = p_\mu dx^\mu$ reads
\begin{equation}
\Theta :=p_\mu dx^\mu |_{\Phi_1 =0=\Phi_2 = \Phi_3}=\eta_kd\xi_k -\kappa ds .
\end{equation}
The unit vector $\xi_n$ can be parametrized by $N-1$ independent coordinates
$\tilde{\theta}^a$ $(a=1,...,N-1)$
and for the orthogonal vector $\eta_n$ one can use $N-1$
additional parameters $\tilde\rho_b$, uniquely defined by
\begin{equation}
\eta_n=\partial_a\xi_n (\tilde\theta) g^{ab}(\tilde\theta)\tilde\rho_b,
\end{equation}
where $g^{ab}(\tilde\theta)$ is the inverse to the induced metric tensor
on the unit sphere $\xi_k\xi_k=1$.  

Neglecting the exact form $\kappa ds$ in (11) 
and using the independent coordinates 
($\tilde{\theta}^a, \tilde{\rho}_a$), we obtain 
the canonical 1-form on $\Gamma_{ph}$ 
\begin{equation}
\Theta =\tilde{\rho}_ad\tilde{\theta}^a.
\end{equation}
Due to the isomorphism between $TS^{N-1}$ and $T^*S^{N-1}$ 
(realized by 
the induced metric on the unit sphere $\xi_k\xi_k=1$),
 the variables $\xi_n$ and $\eta_n$ can be considered as
 functions on $T^*S^{N-1}$. Thus, Eq.(9) defines
 all dynamical integrals (4) 
 as functions on $T^*S^{N-1}$, i.e. 
${M}_{\mu\nu}={M}_{\mu\nu}(\tilde{\rho}, \tilde{\theta})$.
  
Note that the gauge fixing $x^0 -\tau =0$ in the initial action (2)
gives time-dependent non-singular
Lagrangian with the configuration space $S^{N-1}$. 
In the Hamiltonian formulation it leads to the phase-space $T^*S^{N-1}$,
which coincides with our gauge invariant reduction procedure. 

Excluding the variable $\sinh s$, from (10) we find 
\begin{equation}
x_n = \frac{x_0}{\sqrt{\kappa^2+\eta^2}}~\eta_n +
\frac{r\sqrt{\kappa^2+\eta^2 +m^2x_0^2}}{\sqrt{\kappa^2+\eta^2}}~\xi_n,
\end{equation}
which describes the particle trajectories on $\mathbf{H}_N$.
A given point $(\xi_n, \eta_n)$ of $\Gamma_{ph}$ defines the particle
trajectory uniquely. 
Thus, $\Gamma_{ph}$ can be associated with the space
of trajectories as well. According to (14) the `space vector' $x_n$
lays on the plane defined by two orthogonal vectors $\xi_n$ and
$\eta_n$. At the `moment' $x_0=t$ the vector $x_n$ is on the circle of radius 
$\sqrt{x_nx_n} =\sqrt{r^2+t^2}~$. The rotation angle $\alpha(t)$
of the vector $x_n$ from $x_0=0$ to $x_0=t$ is given by
\begin{equation}
\alpha (t) =\arcsin \frac{\eta t}{\sqrt{\kappa^2+\eta^2}~\sqrt{t^2+r^2}}.
\end{equation}
Since $\alpha (t)<\pi /2$, one can speculate that
geodesics of two signals sent in 
different directions never cross. Thus, an `observer' can never see
multiple images of a `cosmic object'.
On the other hand, comparing the `distance' between two nearby geodesics
the observer can detect expansion (for $x_0 >0$) of the `universe'.

\section{Quantization}

The canonical quantization of our reduced system implies realization of 
 commutation relations
\begin{equation}
[\hat\rho_a, \hat{\theta}^b]=-i\hbar\delta_a^b
\end{equation}
and construction of $\hat{M}_{\mu\nu}$ operators by classical expressions
\begin{equation}
\hat{M}_{\mu\nu}:={M}_{\mu\nu}(\hat\rho, \hat{\theta}),
\end{equation}
using definite prescription for the ordering 
of $\hat\rho_a$ and $\hat{\theta}^a$
operators in (17). 
This prescription should lead to a self-adjoint representation
of $so(1,N)$ algebra.
In the case the symmetry generators ${M}_{\mu\nu}$
are linear in momenta $\tilde\rho_a$ the ordering problem can be solved
by a simple symmetrization procedure [1] (see below).

According to (9) and (12) the generators of space rotations
$M_{mn}$ are linear in momenta $\tilde{\rho}_a$,
whereas the boosts $J_n$ are not. In [1]
we have examined the problem of linearization (in momenta)
of symmetry generators
for the case $N=2$. We have found 
transformation to new canonical variables 
which linearize all three generators $M_{01}, M_{02}$
and $M_{12}$. 
In what follows we show that the linearization procedure can be 
generalized for arbitrary $N$ in a covariant form
and we use it for the construction of $\hat{M}_{\mu\nu}$ operators. 

Let us consider transformation to the new 
physical variables $u_n$ and $v_n$ 
\begin{equation}
u_n :=\frac{\kappa}{\sqrt{\kappa^2+\eta^2}}\xi_n +
\frac{1}{\sqrt{\kappa^2+\eta^2}}\eta_n,
~~~~v_n :=-\frac{\eta^2}{\sqrt{\kappa^2+\eta^2}}\xi_n +
\frac{\kappa}{\sqrt{\kappa^2+\eta^2}}\eta_n .
\end{equation}
Transformation (18) is a rotation in ($\xi_n, \eta_n$) plane
by the angle 
\begin{equation}
\beta =\arcsin \frac{\eta}{\sqrt{\kappa^2+\eta^2}}.
\end{equation}
Comparing (19) with (15), we see that Eq.(18) describes transformation
to the asymptotic variables at $t\rightarrow +\infty$.

Making use of (8) and (9) we obtain
\begin{equation}
u_ku_k=1,~~~~~~~~~~~~~~u_kv_k =0,
\end{equation}
\begin{equation}
v^2:=v_kv_k =\eta^2,~~~~~~~~
\eta_kd\xi_k =u_kdv_k-\frac{\kappa dv^2}{2(\kappa^2+v^2)}, 
\end{equation}
and 
\begin{equation}
M_{mn} =u_mv_n -u_nv_m,~~~~~~~~J_n =\kappa u_n-v_n.
\end{equation}

 The variables $u_n$ satisfy the equations
$$
\{\{ {M}_{\mu\nu},u_n\},u_m\}=0~~~~~\mbox{and} ~~~~~\{u_n,u_m\}=0 
$$
(for any $\mu ,\nu, n, m$), which  corresponds to the
choice of polarization in the method of geometric quantization.
In [1] these conditions were used to find the variables
linearizing the dynamical integrals for the case $N=2$.

Due to (20), the variables ($u_n,v_n$) define the manifold
$T^*S^{N-1}$ in the same way as ($\xi_n,\eta_n$). 
We parametrize the unit sphere 
$u_ku_k=1$ by the coordinates $\theta^a$ 
(without tilde) and again introduce the coordinates $\rho_a$ (without tilde) 
defined by (see (12))
\begin{equation}
v_n = \partial_au_n(\theta )g^{ab}(\theta )\rho_b,
\end{equation}
where now $g^{ab}(\theta )$ is the inverse to
\begin{equation}
g_{ab}(\theta):=\partial_au_k(\theta)
\partial_bu_k(\theta). 
\end{equation}
In the new coordinates $(\rho_a,\theta^a)$ we have again
 the canonical symplectic form  $\sigma :=d\Theta =d\rho_a\wedge d\theta^a$
(see (13) and (21)).
 
 \vspace{0.5cm}
\noindent
It is natural to choose $L^2(S^{N-1})$ as a Hilbert space,
with the scalar product 
\begin{equation}
\langle\Psi_2|\Psi_1\rangle :=\int d\theta\sqrt{g(\theta )}
~\Psi_2^*(\theta )\Psi_1(\theta ), 
\end{equation}
where $g(\theta ):=det g_{ab}(\theta)$ and $d\theta\sqrt{g(\theta )}$
is the invariant measure on $S^{N-1}$.

The Hermitian operators $\hat\rho_a$, which satisfy (16), have the form
\begin{equation}
\hat\rho_a =-i\hbar\partial_a -{i\hbar}\frac{\partial_a g}{4g}. 
\end{equation}
All generators $M_{\mu\nu}$ are now linear in $\rho_a$
coordinates
 \begin{equation}
{M}_{\mu\nu}({\rho}, {\theta}) =
{\rho}_aA^a_{\mu\nu}(\theta )+B_{\mu\nu}(\theta ),
\end{equation}
where $B_{\mu\nu}(\theta )$ are functions  and
$A^a_{\mu\nu}(\theta )$ are the components of vector fields on $S^{N-1}$.
For the corresponding operators (17)
we apply the following symmetrization prescription
\begin{equation}
\rho_aA^a_{\mu\nu}(\theta )
\rightarrow \frac{1}{2}[\hat{\rho}_aA^a_{\mu\nu}(\theta ) +
A^a_{\mu\nu}(\theta )\hat{\rho}_a]. 
\end{equation}
The prescription (28) provides Hermiticity of $\hat{M}_{\mu\nu}$
operators and preserves the classical commutation relations 
of the dynamical integrals [1]. 

Making use of (26) and (28) we obtain
\begin{equation}
\hat M_{\mu\nu} =-i\hbar [A_{\mu\nu} +B_{\mu\nu}(\theta )] 
-\frac{i\hbar}{2}\nabla (A_{\mu\nu}),
\end{equation}
where $A_{\mu\nu}$ is a vector field on $S^{N-1}$
\begin{equation}
A_{\mu\nu} := A^a_{\mu\nu}(\theta )\partial_a ,
\end{equation}
\begin{equation}
\nabla (A_{\mu\nu}) :=\partial_aA^a_{\mu\nu}(\theta ) +
\frac{\partial_ag(\theta)}{2g(\theta)}A^a_{\mu\nu}(\theta )
=\nabla_aA^a_{\mu\nu},
\end{equation}
and $\nabla_a$ is the operator of covariant derivative on the Riemannian
manifold [8].
Thus, the operators $ \hat{M}_{\mu\nu}$
do not depend on the choice of 
$\theta^a$ coordinates. 
 
To calculate the term $\nabla (A_{\mu\nu})$ we introduce the vector fields 
$Y_n~
(n=1,...,N)$ defined as solutions of the equations
\begin{equation}
G(Y_n,\cdot~)=du_n(\cdot ),
\end{equation}
where $G:=g_{ab} d\theta^a d\theta^b$ is the symmetric 2-form 
of the induced metric on $S^{N-1}$.
Due to nondegeneracy of $G$, Eq.(32) has unique solution for each $n$ 
 given by
\begin{equation}
Y_n=Y_n^a(\theta)\partial_a,~~~~~~~\mbox{where}~~~~~~~
Y_n^a(\theta) = g^{ab}(\theta )\partial_bu_n(\theta ).
\end{equation}

The vector fields $Y_n$ and the functions $u_n$ satisfy 
the following relations (see Appendix)
\begin{equation}
Y_n(u_m)= \delta_{mn}-u_mu_n ,
\end{equation}
\begin{equation}
\nabla (Y_n) =-(N-1)u_n .
\end{equation}

According to (22), (23) and (27) the vector fields $A_{\mu\nu}$ 
associated with
the space rotations and the boosts are
$A_{mn} =u_mY_n-u_nY_m$ and $A_{n0} =-Y_n$  
respectively; the functions $B_{\mu\nu}$
read: $B_{mn}=0$ and $B_{n0}=\kappa u_n$.
Due to (31), (34) and (35) we have 
 $$\nabla (u_mY_n) =\delta_{mn} -Nu_mu_n ,$$
which leads to $\nabla (A_{mn}) =0$. As a result we obtain
\begin{equation}
\hat M_{mn} =-i\hbar (u_m Y_n -u_nY_m),~~~~~~~~ \hat J_n =
\left (\kappa -\frac{i\hbar}{2}(N-1)\right ) u_n +i\hbar Y_n. 
\end{equation}
The operators (36) realize the commutation relations of $so(1,N)$ algebra,
which follows from (34) and (A.6).

Taking into account that $u_nY_n=0$, we find the quantum Casimir
number corresponding to (6)
\begin{equation}
 \hat C=\frac{1}{2}\hat M_{\mu\nu}\hat M^{\mu\nu}=\kappa^2+
 \frac{\hbar^2(N-1)^2}{4}.
\end{equation}

Let us specify the representation of $SO_\uparrow (1,N)$ group
corresponding to (36). 
For any $\Lambda^\mu\,_\nu\in SO_\uparrow (1,N)$ one has the decomposition
 $\Lambda^\mu\,_\nu =R^\mu\,_\beta\, B^\beta\,_\nu$, 
where $R^\mu\,_\nu$ is the space rotation
\begin{equation}
 R^0\,_0=1,~~~R^0\,_n=0=R^n\,_0,~~~ 
R^m\,_n=\Lambda^m\,_n -\frac{\Lambda^m\,_0\Lambda^0\,_n}{1+\Lambda^0\,_0}
\end{equation}
and $B^\mu\,_\nu$ is the boost
\begin{equation}
B^0\,_0=\Lambda^0\,_0,~~~B^0\,_n=\Lambda^0\,_n=B^n\,_0,~~~
B^m\,_n=\delta^m\,_n+\frac{\Lambda^0\,_m\,\Lambda^0\,_n}{1+\Lambda^0\,_0}. 
\end{equation}
The boost (39) 
 is characterized by the transformation parameter $\sigma$
and the unit vector $\vec\zeta :=(\zeta_1,...,\zeta_N)$ 
defined by
\begin{equation}
\cosh\sigma =\Lambda^0\,_0,~~~~~~~ 
\zeta_n \sinh \sigma= \Lambda^0\,_n~~~~( \sigma >0).
\end{equation} 

One can prove (see Appendix) that
the action of the boost operator  
\begin{equation}
 \hat B(\sigma ,\vec\zeta) :=\exp \left(-\frac{i}{\hbar} 
 \sigma\zeta_n\hat J_n\right )
\end{equation}
on the wave function $\Psi\in L^2(S^{N-1})$ has the form
\begin{equation}
 \hat B(\sigma,\vec\zeta )\Psi(\vec w) =
 [\cosh\sigma+\langle\vec w\cdot\vec\zeta\rangle
 \sinh\sigma]^{-z}~ 
 \Psi\left (\vec F (\vec w;\sigma,\vec\zeta)\right ) ,
\end{equation}
where 
$$
\vec w:=(w_1,...,w_N)\in S^{N-1}~~~~~~(~ u_n(\vec w)=w_n ~),
$$
$$
\langle\vec w\cdot\vec\zeta\rangle :=w_n\zeta_n,~~~~~~~~
z:=\frac{N-1}{2}+i\frac{\kappa}{\hbar},
 $$
 and 
$\vec F(\vec w;\sigma,\vec\zeta)$ is the flow generated by the vector 
field $Y_{\vec\zeta}:=\zeta_nY_n$. This flow is given by (see Appendix)
\begin{equation}
\vec F(\vec w;\sigma,\vec\zeta):=\frac{\vec w +[\sinh \sigma +
 \langle\vec w\cdot\vec\zeta\rangle (\cosh \sigma -1)]\vec\zeta}
 {\cosh\sigma+\langle\vec w\cdot\vec\zeta\rangle
 \sinh\sigma}. 
\end{equation}
Since the operators $\hat M_{mn}$ are vector fields,
the action of the rotation operator on the wave function $\Psi (w_n)$
transforms only its argument 
\begin{equation}
w_n\rightarrow w_mR^m\,n.
\end{equation}

Making use of (38)-(44) we find the transformation  
corresponding to $\Lambda^\mu\,_\nu$ 
\begin{equation}
\Psi (w_n)\rightarrow [\Lambda^0\,_0+\Lambda^k\,_0w_k]^{-z}~
\Psi \left(\frac{\Lambda^m\,_nw_m
+\Lambda^0\,_n}{\Lambda^0\,_0+\Lambda^l\,_0w_l}\right ) ,
\end{equation}
which defines the unitary irreducible representation of $SO_\uparrow (1,N)$
group. For the case $N=2$ it is equivalent to the representation of
$SL(2,R)$ group [9].

\section{Remarks}

It should be noted that the prescription (28) for the operator 
ordering is not unique. One can show that Hermiticity of the symmetry 
generators could be achieved by the more general prescription
$$
M_{\mu\nu}\rightarrow \hat{\tilde{M}}_{\mu\nu}:=
\hat{M}_{\mu\nu} +c\nabla(A_{\mu\nu}),
$$
where $\hat{M}_{\mu\nu}$ is given by (29) and $c$ is any real number.
This freedom and related details of the representation (45)
 will be discussed elsewhere.

In the case $N=4$, the induced metric tensor on the hyperboloid
satisfies the Einstein equation
$$R_{\mu\nu}-\frac{1}{2}Rg_{\mu\nu} +\Lambda g_{\mu\nu} =0,$$
with the cosmological term $\Lambda =3/r^{2}$.
Therefore, it is de Sitter's type metric and 
the corresponding spacetime can be considered as a `toy model'
of the Universe.

As it was mentioned in Introduction (discussing the case $N=2$),
there exists the relationship between global properties (topology, symmetry)
of spacetime and physical phase-space.  The corresponding quantum 
system should take into account this relationship.
As we show the correlation among spacetime, physical phase-space
and the corresponding quantum system 
can be generalized to the case of $N$ dimensional
 hyperboloid.
Due to this correlation one can consider the inverse problem:
finding spacetime(s), which leads to a given physical phase-space 
and corresponding quantum system. Such a problem could be interesting in the
context of cosmology.

Massive scalar particle is the simplest physical object. As we have shown,
its dynamics (both classical and quantum) can be described completely,
as in the case of the flat Minkowski space. 
It would be interesting to investigate 
the dynamics of other systems like spinning particle, (super)string, etc.

\vspace{0.3cm}

{\bf Acknowledgments   }

\vspace{0.3cm}

This work was supported by the grants from:
INTAS  (96-0482), RFBR,
the Georgian Academy of Sciences, the Polish Academy of Sciences and the
So{\l}tan Institute for Nuclear Studies.

\setcounter{equation}{0}
\def\theequation{A.\arabic{equation}}

\section{Appendix}

 According to (33) one has 
\begin{equation}
Y_n(u_m)=g^{ab}\partial_au_n\partial_bu_m ,
\end{equation}
which does not depend on the choice of local coordinates $\theta^a$
on $S^{N-1}$.
We choose $\theta^a=u_a,~ a=1,...,N-1$. The corresponding 
metric tensor (24)
takes the form
\begin{equation}
g_{ab} =\delta_{ab} +\frac{u_au_b}{u_N^2},~~~~~~~
\mbox{where}~~~~ u_N^2 =1-\sum_{a=1}^{N-1}u^2_a.
\end{equation}
The inverse to the matrix (A.2) is 
\begin{equation}
g^{ab} =\delta_{ab} -u_au_b,
\end{equation}
and (A.1) leads to (34), for any $n,m\in \{1,...,N\}$.

The $N-1$ dimensional `vector' $u_a~ (a=1,...,N-1)$ is an eigenvector of the
matrix (A.2) with the eigenvalue $1/u_N^2$. Other $N-2$ eigenvectors 
are orthogonal to $u_a$ and have eigenvalues equal to one. Thus, 
the determinant $g$ of the matrix (A.2) reads  
\begin{equation}
g :=det g_{ab} ={1}/{u_N^2}.
\end{equation}
By (A.2) and (A.4), for any $n\in \{1,...,N\}$, we obtain
\begin{equation}
\nabla (Y_n) :=\partial_a(g^{ab}\partial_bu_n) +
\frac{\partial_ag}{2g}g^{ab}\partial_bu_n =-(N-1)u_n.
\end{equation}

\vspace{0.3cm}
The commutator of vector fields $Y_m$ and $Y_n$ has the form
\begin{equation}
Y_mY_n-Y_nY_m =u_mY_n-u_nY_m. 
\end{equation}
The validity of (A.6) results from the fact that the action of both sides
 of (A.6) on $u_k$, for any $k \in \{1,...,N\}$, gives the same result:
 $u_m\delta_{nk}-u_n\delta_{mk}$.
Since the set of functions $u_k$ forms
 (over)complete set on $S^{N-1}$, one has (A.6).

\vspace{0.3cm}

The flow $\vec{F} :=\vec{F}(\vec w;\sigma,\zeta)$ corresponding to
the vector field $Y_{\vec\zeta}=\zeta_nY_n$ is defined by (see (34))
 \begin{equation}
\partial_\sigma\vec{F} =\vec\zeta-\langle \vec{F}\cdot
\vec\zeta\rangle\vec{F}, 
\end{equation}
 and
 \begin{equation}
 \vec{F}(\vec w;0 ,\vec\zeta)=\vec w,
 \end{equation}
 where (A.8) is the initial value for (A.7).
 
 Multiplying (A.7) by the vector $\vec\zeta$, we get
 \begin{equation}
\partial_\sigma\langle\vec{F}\cdot\vec\zeta\rangle =
1-\langle \vec{F}\cdot\vec\zeta\rangle^2. 
\end{equation}
Solution to (A.8)-(A.9) reads
 \begin{equation}
\langle\vec{F}\cdot\vec\zeta\rangle =               
\frac{\sinh\sigma +\langle\vec w\cdot\vec\zeta\rangle \cosh\sigma}
{\cosh\sigma +\langle\vec w\cdot\vec\zeta\rangle\sinh\sigma }.                                                  
\end{equation}                                                                                                                                                   
Substitution of (A.10) into (A.7) gives the linear differential equations
for $\vec F$. Integration of these equations leads to (40).

Taking into account the form of $\hat J_n$ operators, from (36) and
(38) we have
\begin{equation}
 \hat B(\sigma ,\vec\zeta)\Psi(\vec w) =f(\vec w;\sigma ,\vec\zeta)
 \Psi\left (\vec F
 (\vec w;\sigma ,\vec\zeta)\right ),
\end{equation}
where $f$ is some function. Since the right hand side of (A.11)
should define one parameter group (with respect to $\sigma$) and
$f(\vec w; 0,\vec\zeta) =1$, we find  
$$f(\vec w;\sigma ,\vec\zeta)=
[\cosh\sigma+\langle\vec w\cdot\vec\zeta\rangle\sinh\sigma ]^{-z}.$$

\end{document}